# Role of surface chemistry on incorporation of nanoparticles within calcite single crystals

Giulia Magnabosco[#], Iryna Polishchuk[§], Francesco Palomba[#], Enrico Rampazzo[#], Luca Prodi[#], Joanna Aizenberg[‡], Boaz Pokroy[§, *], and Giuseppe Falini[#, *]

[‡] Dipartimento di Chimica "Giacomo Ciamician", Alma Mater Studiorum-Università di Bologna, Via F. Selmi 2, 40126 Bologna, Italy

[§] Department of Material Sciences and Engineering and the Russel Berrie Nanotechnology Institute Technion-Israel Institute of Technology 32000 Haifa, Israel.

[‡] John A. Paulson School of Engineering and Applied Sciences, Department of Chemistry and Chemical Biology, Wyss Institute for Biologically Inspired Engineering, Harvard University, Cambridge, MA 02138, USA.

ABSTRACT: Inclusion of additives into calcite crystals allows one to embed non-native proprieties into the inorganic matrix and obtain new functional materials. Up to now, few parameters have been taken into account to evaluate the efficiency of inclusion of an additive. Taking inspiration from Nature, we grew calcite crystals in the presence of fluorescent silica nanoparticles carrying different functional groups (PluS-X) to investigate the effect of surface chemistry on the inclusion of the additives. PluS-X allowed us to keep constant all the particle characteristics, including size, while changing exposed functional groups and thus Zeta-potential. The effect on crystal morphology, the loading and distribution of PluS-X within the crystals have been evaluated with different microscopy techniques. Our data indicate that hydroxyl functionalized particles are entrapped more efficiently inside calcite single crystals without distortion of the crystal structure and inhibition of the growth.

INTRODUCTION

Biomineralization has always been fascinating for scientists, who first tried to understand the strategies that allow living organisms to control crystal growth[1-4] and then to imitate these principles and exploit them in the synthesis of new materials. [5-7] The key to the incredible control over crystalline morphology and polymorphism found in Nature is the organic matter present during crystal nucleation and growth. [3,8-12] Calcium carbonate ($CaCO_3$) is among the most studied systems since it is one of the most abundant materials produced both biogenically and geologically. Outstanding examples of biogenic $CaCO_3$ are sea urchin spines, coccolithophores and nacre. All these biological systems are shaped by a small quantity of organic phase.[9,13]

Organic compounds interact with the inorganic phase in two different ways: they can act as a surface over which the inorganic phase forms or they can adsorb onto the surface of the growing crystals, thus hindering the growth of the crystal in certain crystallographic directions. [8,14-17] Both principles are exploited in biomineralization, where crystals usually grow inside a preformed scaffold that acts as a template while soluble additives modify the morphology of the crystals. Great attention was paid to the composition of the templating surfaces and it has been found that they are mainly composed of proteins rich in negatively charged residues.[18,19] In particular, *in vitro* studies demonstrated that self-assembled monolayers (SAM) exposing carboxylated and sulfonated residues are able to coordinate calcium ions and favor the appearance of (015) and (001) faces of calcite, respectively. [20] Opposed to this, when proteins are present in solution they can selectively adsorb onto certain crystal faces inhibiting their growth, therefore modifying the final crystal morphology and eventually get incorporated within the lattice. [21]

The inclusion mechanism of different classes of molecules, ranging from amino acids, [22,23] to drugs, [24,25] micelles and nanoparticles,[26-31] within calcite single crystals has been extensively studied. These researches show that it[BP1] is difficult to ascribe the inclusion of a molecule by only considering the chemistry of functional groups[34,35] since their modification can change the chemical, physical and sterical features of the whole molecule. An alternative approach is the use of nanoparticles, in which we can change the surface functional groups while keeping the bulk features, as the morphology, unchanged. However, to our knowledge to date, only a limited number of studies[32,33] have related the yield of the nanoparticle occlusion to its surface chemistry. Various nanoparticles have been shown to successfully interact with several minerals.[36] The crucial role played by surface chemistry was demonstrated by the need of a carboxylic corona on polystyrene nanoparticles to obtain templated calcite crystals via ACC.[37] Conghua et al. showed that latex nanoparticles can template the surface of calcite crystals due to the interaction of the exposed carboxylic groups with the nucleating crystals. Despite that, no incorporation of the particles inside the bulk of the crystal was obtained. When the carboxilyic terminal group was substituted by another negatively charged group, no interaction was detected.[38] Kim et al. showed that the interaction between the particles and the growing crystals can be ascribed to the nature of the particle surface rather than to its electrostatic charge.[7,39] Recently, the same research group has shown that an optimal ratio between hydroxyl and carboxylate groups is required for an efficient entrapment of functionalized copolymer nanoparticles within calcite crystals; moreover an increase of density of carboxylate groups does not promote the embedding of the particles as expected. [7,26] The relevance of carboxylate groups, with respect to other functional groups, has been confirmed in a recent study on a series of silica-loaded tetrablock copolymer vesicles with constant dimensions. This material led to[BP2] the discovery that the degree of polymerization of the anionic polymer plays a decisive role in dictating the extent of occlusion within calcite crystals, which is favored by carboxylate relative to other negatively charged functional groups.[40]

In this research we propose silica nanoparticles – synthesised using a micelle-assisted strategy – as additives to investigate the role of surface chemistry in the entrapment within calcite crystals. These nanoparticles are composed of a core that is not influenced by the environment and by a shell that can carry, differently form previous studies, a great variety of anionic, cationic or neutral functional groups.[41] Furthermore, their chemistry is not influenced by the pH of the solution as strongly, since it[BP3] occurs in micelles and block copolymers, thus allowing the investigation of a variety of crystal growth conditions in which other species would not be stable.[42]

RESULTS AND DISCUSSION

In this work, we used fluorescent silica nanoparticles functionalized with a polyethylene glycol shell (PluS-X, Figures S1-3) that have been synthesized for application in nanoscience due to their easy detection even in complex matrices.[42,43] PluS-X (Figure 1a) are 25 nm highly monodisperse nanoparticles made of a 10 nm rigid silica core and a soft Pluronic shell (PEGylated polymer) terminated with hydroxyl groups (PluS-OH). The nature and the amount of the terminal groups present on the Pluronic shell can be easily modified to expose different functions (protocol in SI, Table S1), and to tune the surface charge properties of PluS-X. In particular, together with PluS-OH that do not present charged groups at their periphery, we synthesized PluS-NH$_2$ and PluS-COOH, presenting the same number of amine and carboxylic groups, respectively, in order to investigate the effect of these surface functional groups on the inclusion of the additives within calcite single crystals. This approach offers a further advantage of minimizing the effect of other otherwise important characteristics of the additive by keeping constant PluS-X size (core size distribution determined by TEM and DLS-determined hydrodynamic radius in figures 1b-c, S4 and S5), and the number of functional groups per surface, allowing us to directly relate the effect of PluS-X on the calcite crystal.

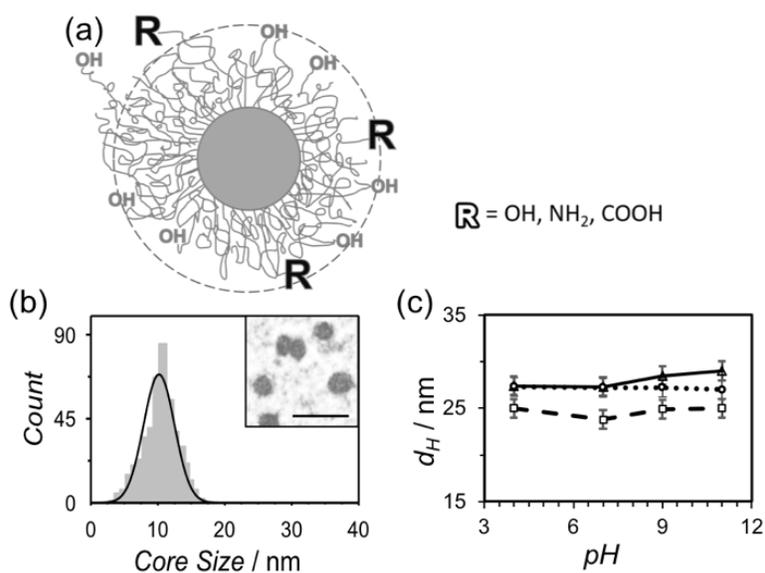

**Figure 1**. (a) Schematic illustration of PluS-X structure, (b) distribution of silica core diameter and transmission electron microscopy (TEM) image of PluS-X silica core (scale bar 20 nm). (c) trends of the hydrodynamic diameter of PluS-OH (dashed line), PluS-NH$_2$ (dotted line), PluS-COOH (solid line), at different pH (CaCl$_2$ 10 mM, [PluS-X]= 1 μM).

The low density of charged functional groups agrees with the measured zeta-potential values, in contrast to the high values reported in literature data.[44]

As mentioned in the experimental section, calcite single crystals were grown in the presence of various concentrations of PluS-X using the vapor diffusion technique. Crystals formed in all the examined conditions and no relevant change in the number of precipitated crystals was detected, meaning that PluS-X does not inhibit the nucleation phase. Furthermore, only single crystals were observed, except when the highest concentration of PluS-NH$_2$ was used, indicating the absence of secondary nucleation events.

Fluorescent dyes, which allow a convenient localization and quantification of the PluS-X, are chemically incorporated into the core and protected from the environment by the silica core matrix and by the shell. This prevents any influence of the external environment on the absorption and emission properties of the dye. Exploiting the fluorescence of the PluS-X, it was possible to quantify their embedding in the crystalline lattice (Figure 2).

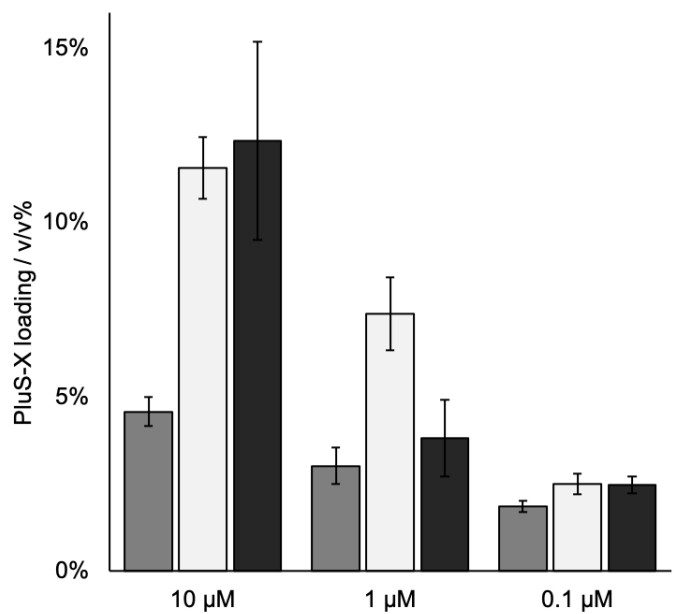

**Figure 2.** PluS-X loading (v/v %) in calcite crystals grown in the presence of PluS-OH (white), PluS-NH$_2$ (black), PluS-COOH (gray). (p-value between 1 μM PluS-NH$_2$ and 1μM Plus-OH=0.02; p-value between 1 μM PluS-COOH and 1μM Plus-OH = 0.006). The hydrodynamic

diameters of the PluS-X used for calculation is 25 nm. PluS-X loading in calcite crystals as (v/v %) considering only the silica core and as wt.% are reported in figure S6 and S7, respectively.

Each PluS-X has been functionalized with an average value of 10 Rhodamine B (RB) dyes yielding very bright nanoarchitectures (photoluminescence quantum yield $\Phi_{PL}$ = 0.24). The concentration of the PluS-X embedded in the crystal lattice has been measured by recording the fluorescence of the RB in solution upon the dissolution of 3 independent sets of crystals grown simultaneously in the same conditions for each PluS-X, while the concentration of $Ca^{2+}$ has been measured by atomic absorption allowing the computation of PluS-X content as the ratio of the PluS-X divided by the $Ca^{2+}$ concentration (Figure 2). Although the PluS-X content entrapped within the calcite crystals is low, the high quantum yield of PluS-X allows an accurate detection and a comparative study among differently functionalized PluS-X.

At low concentration, PluS-X content is comparable for all the samples. At intermediate concentrations, we can observe that PluS-OH is more likely to be incorporated than PluS-NH$_2$ and PluS-COOH, while at highest concentration the amount of PluS-NH$_2$ loaded is comparable to that of PluS-OH. The latter effect can be ascribed to the formation of polycrystalline aggregates when the crystallization process is carried out in the presence of 10 µM PluS-NH$_2$. This makes it possible for the PluS-NH$_2$ to be entrapped between crystalline domains, as confirmed by confocal microscopy images that show a build up of PluS-NH$_2$ at the interface of different crystalline domains when looking at inside sections of crystals (figure S10). Furthermore, since in the experimental conditions the growing calcite crystals have a negatively charged surface[45,46] it discourages the negatively charged PluS-COOH from adsorbing onto it (Figure S4). [7,26]

With the aim of better understanding the inclusion process of PluS-X within the calcite lattice, the amount of PluS-X desorbed by the crystals has been determined by measuring the amount of PluS-X released in 750 mL of a saturated $CaCO_3$ solution in which the crystals were submerged for 72 hours under shaking. These crystals were carefully washed before the measurements. The amount of PluS-X released in solution by the crystals was not detectable by fluorescence measurements, except for the crystals with the highest content of PluS-NH$_2$ which released less that 0.1 % of bulk content.

The change in morphology of the final crystals is related to the adsorption of the PluS-X on the crystalline faces, thus giving information on the mechanism of interaction. Looking at scanning electron microscopy (SEM) pictures (Figure 3) of crystals grown in the presence of different concentrations of PluS-X, we can see that at low concentration of PluS-COOH and PluS-OH, nanoparticles cause the crystal to develop a hopper-like morphology and all PluS-X create holes in the {104} faces. In constrast to the previously cited studies, in which holes on the surface exhibited similar geometry and dimension as compared to the additive used, we observe pores with significantly greater dimensions than those of PluS-X used. The formation of these pores can be ascribed to the fact that the adsorbed PluS-X onto the growing calcite surfaces are objects showing very large sizes with respect to the surface lattice parameters of calcite (a few Å). They behave like an inclusion that can adhere to the growing calcite surface because the adhesion force overcomes the disjoining pressure existing between the object and the crystal. Thus, the advancing calcite growth layers can encompass and englobe the objects that finally will be absorbed/incorporated into the crystal bulk. However, before the incorporation, the flatness of the {10.4} rhombohedra is strongly perturbed by the presence of these foreign objects which change (hydrodynamic effects) the local supersaturation and, consequently, introduce morphological instability to the growing crystal surfaces (small cavities) and rounded edges. The latter phenomenum has been described as due to non specific interactions among interacting objects and calcite crystals, according to the studies reported by Addadi et al.[8, 21]

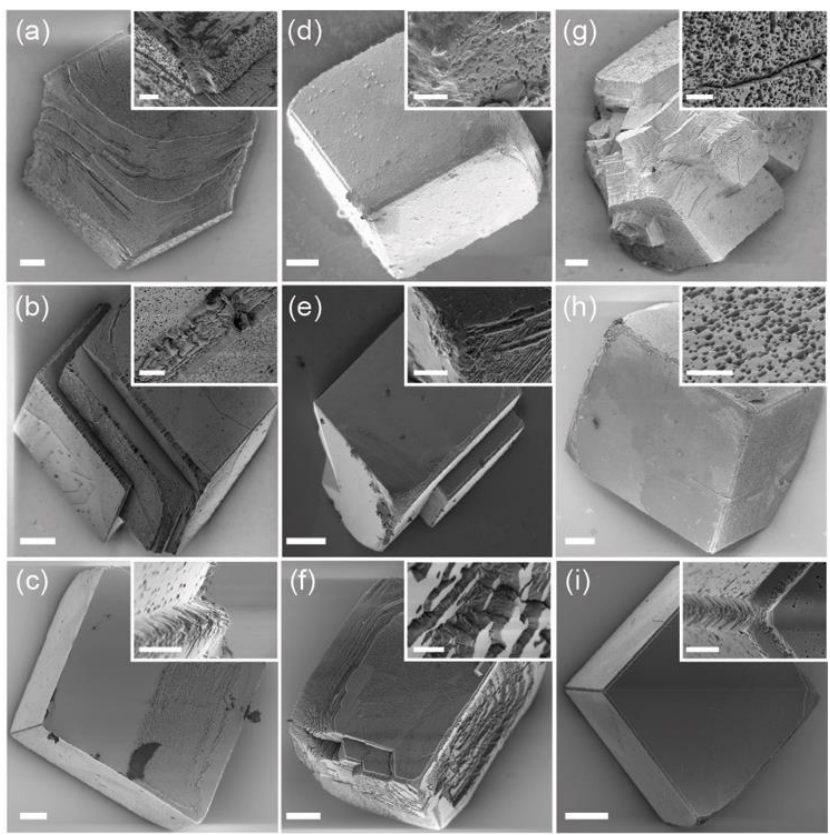

**Figure 3.** SEM micrographs of calcite crystals grown in the presence of (a) 10 µM, (b) 1 µM, (c) 0.1 µM PluS-COOH, (d) 10 µM, (e) 1 µM, (f) 0.1 µM PluS-OH, (g) 10 µM, (h) 1 µM, (i) 0.1 µM PluS-NH$_2$. Scalebar is 10 µm in the main image and 1 µm in the inset. Low magnification images are reported in Figure S9. These images are representative of the population of crystals.

Regarding PluS-NH$_2$, a higher concentration is needed to observe a comparable change in morphology. In fact, at the lowest concentration studied, we observe only a limited quantity of defects on the surface of the crystals while, increasing the concentration of PluS-NH$_2$ to 1 µM, the number of holes increases as the morphology of the crystal remains unaffected. When crystals are grown in the presence of 10 µM PluS-NH$_2$, we observed a strong modification of the morphology and the formation of polycrystalline aggregates, which can be ascribed to the secondary nucleation processes. The formation of polycrystalline aggregations occurs when an additive partially poisons the crystal surface growing sites. As a consequence, an increase of supersaturation occurs and this favors heterogeneous nucleation and secondary nucleation effects. Our experimental setup does not allow us to follow the evolution of the supersaturation during the precipitation process, similarly to the majority of the cited literature. Thus we do not have a direct proof of the mechanism of polycrystalline material formation. We can only suppose that the NH$_2$-fuctionalized particles affect the local supersaturation in the proximity of the crystal surface causing the formation of polycrystalline aggregates.

We hypothesize that PluS-OH - not carrying charged groups at their surface - are better entrapped within the crystal since they do interact with the crystal surface, yet without preventing further growth.[35] This is confirmed by the rhombohedral shape of the crystal grown in the presence of 10 µM PluS-OH.

Unfortunately, we were not able to measure the concentration of the nano-particles on the rough surface. We measured the amount of adsorbed PluS-X that was[BP4] released from the surface after 72 hours of soaking in a saturated solution of calcite, which not necessarily represents the concentration on the surface (Table S2). To verify the incorporation of the particles inside the crystals, we took advantage of the fluorescent dyes linked to the core of the particles that allows us to localize them inside the crystals using confocal microscopy. This avoids the use of techniques, which give information only on small and sections of a limited number of crystals (e.g. FIB-SEM).

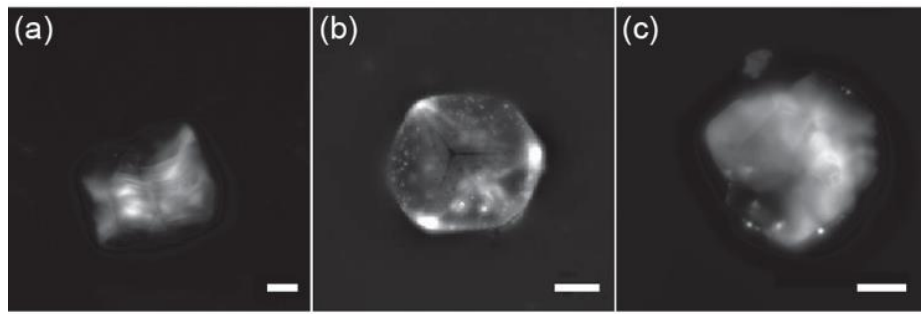

**Figure 4**. Z-stack confocal images of calcite crystals grown in the presence of 10 µM (a) PluS-COOH, (b) PluS-OH and (c) PluS-NH$_2$. Slices are reported in Figures S10-12. Scale bar: 25 µm.

As can be seen in Figure 4, all the particles are embedded inside the crystals rather than just adsorbed on the surface.

According to the high resolution X-ray diffraction analysis, lattice parameters are not significantly affected by the presence of PluS-X (Table S3, Figures S13-15).

To better understand the interaction of PluS-X with the growing calcite surface, we performed overgrowth experiments using pristine calcite crystals as seeds and a reference for the new inorganic phase. In these conditions, no secondary nucleation was observed for any of the samples. We performed atomic scale analysis on the thin sections of the overgrowth crystals obtained in the presence of PluS-OH particles (Figure 5). Using scanning TEM (STEM) coupled to the high-angular annular dark field (HAADF) imaging mode and to an advanced energy dispersive X-ray (EDS) detector[BP5], allowed us to measure the location of PluS-X particles within the overgrown layer with atomic resolution. Our observations suggest that PluS-X localize at the interface between the reference core crystal and the overgrown layer (Figures 5a, b and c). Images obtained in transmission electron microscopy (TEM) mode confirmed that the lattice of the overgrown calcite layer is not disrupted and its further growth is not affected by occluded PluS-X particles (Figures 5d and e). Fast Fourier transforms (FFT) applied to the lattice images acquired from the interface area demonstrate a single-crystalline nature of the crystal, with no distortion of the lattice parameters (Figure 5e, inset). These results fit with previous observations regarding the inclusion of micelles within the calcite lattice, where their embedding of the particle did not produce any perturbation of crystalline order.[29]

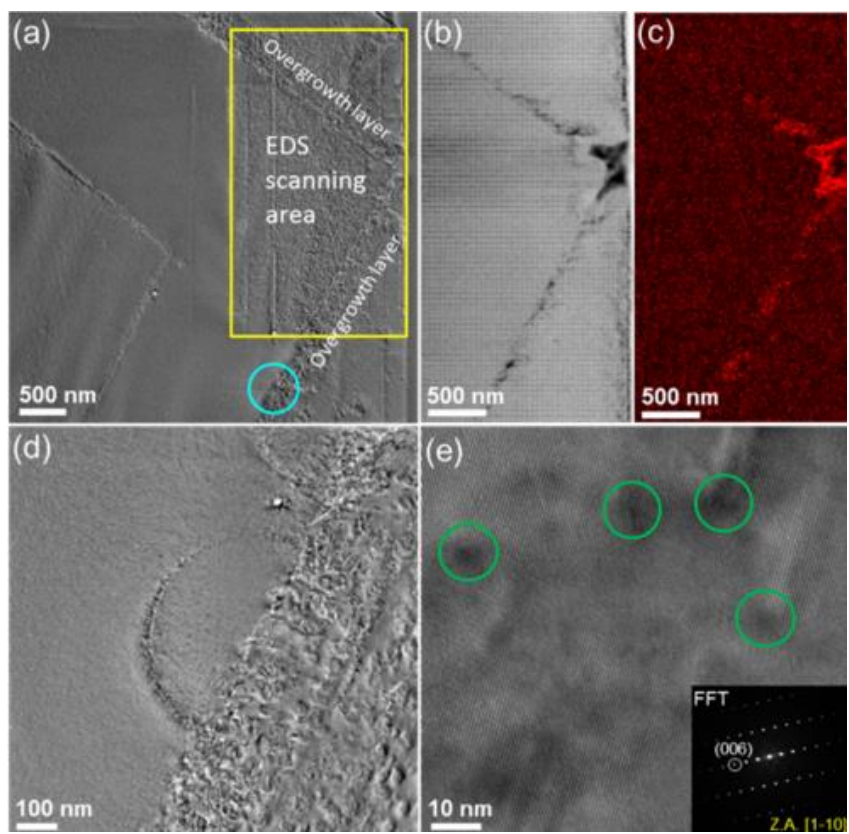

**Figure 5.** High resolution TEM analysis of the overgrowth crystals. (a) STEM image of the analyzed thin section. (b) STEM-HADAAF image of the area marked in yellow in (a). Dark spots correspond to the embedded PluS-X particles due to their lower Z-values as compared to that of the calcite surrounding matrix. (c) STEM-EDS map of the Si Kα peaks revealing localization of the PluS-X. (d) Low magnification TEM image of the area marked in blue in (a). (e) High magnification TEM image demonstrating not disrupted calcite lattice. Some PluS-X are circled in green. Inset – FFT-generated diffraction pattern that can be well fitted to that of a single crystal of calcite.

CONCLUSIONS

In conclusion, we proved that surface chemistry is an important parameter affecting the inclusion of nanoparticles inside calcite single crystals. Our experiments showed that PluS-OH are the most efficiently embedded particles among the ones studied. This is due to the ability of the OH group to adsorb onto the growing surface while not carrying any charge that can cause repulsion in crystallization conditions. PluS-COOH are incorporated in lower quantity due to their negative charge, while PluS-NH$_2$ cause secondary nucleation processes and localize between crystalline domains. TEM analysis performed on thin sections of the overgrown crystal confirms that PluS-X adsorb on the growing calcite surface and do not prevent the growth of the inorganic single-crystalline phase. In addition, we observed that the nanoparticles' inclusion does not induce a distortion of the crystal lattice.

These observations are relevant in material science and in biomineralization. Regarding the former, enhancing the inclusion of PluS-X into calcite single crystals exploiting the shell chemistry allows one to integrate non-native functions into the material, with the advantage of integrating multiple functionality, within the core of the particles and having[BP6] no effect on the inclusion rate. This possibility could be applied to drug delivery, combining the passive delivery of PluS-X in the acidic tumor environment due to dissolution of CaCO$_3$ together with the active delivery that can be obtained functionalizing PluS-X with specific cancer targeting agents and anticancer drugs. In the latter, biomineralization, the occlusion of nanoparticles into a crystalline framework is a widespread phenomenon, which often involves macromolecular ultra-structures containing polysaccharides and thus exposing hydroxyl groups, with minimal or no distortion of the crystal lattice.

EXPERIMENTAL SECTION

*Synthesis of Pluronic Silica Nanoparticles* Core shell silica-PEG nanoparticles bearing carboxyl and amine groups were synthesized as previously reported[43] and described in SI (see also Figures S1-3). In the synthetic procedure a total amount of 200 mg of Pluronic F-127 surfactants was used (Table S1). The surfactant(s) and the alkoxysilane derivative of rhodamine B dye (1.2 mg, 0.0016 mmol) were solubilized with 1-2 mL of dichloromethane in a scintillation vial. The organic solvent was removed under reduced pressure and 3.2 mL of acetic acid 1 M were added. After the complete dissolution of the surfactant and of the dye the silica core was formed by the addition of 360 μL of TEOS (1.60 mmol). The reaction was conducted under stirring for 3 hours, after that 20 μL of TMSCl were added as end-capping agent. Overnight the solution was dialyzed against water with a regenerated cellulose membrane (MWCO 12 kDa).

*Nanoparticles quantification* MALS-RI and AF4-MALS-RI measurements (MALS: Multi angle Light scattering) were used to measure the molecular weight of the PluS-OH nanoparticles in buffered water suspensions. This allowed to calculate the concentration of the nanoparticles obtained by this synthetic method. The PluS-OH NP concentration has been determined even using fluorescent titrations. The excitation energy conferred to rhodamine doped PluS-OH was almost quantitatively transferred to Cy5 dyes hosted in the PEGylated shell of the nanoparticles. [47] The possibility to determine the PluS-OH NPs concentration by fluorimetric titration is linked to the extremely high efficiency (close to 100%) by which excitation energy can be transferred between donors and acceptors fluorescent dyes embedded in the same nanoparticle.[48-50] The results that we obtained by the two different methods were in close agreement.

*Nanoparticles hydrodynamic diameter distributions and Zeta potential* These data were recorded by Dynamic Light Scattering measurements with a Malvern Nano ZS instrument equipped with a 633 nm laser diode. Samples were filtered with 0.45 µm RC filters and then housed in disposable polystyrene cuvettes of 1 cm optical path length, using [PluS-X] = 1 mM dispersed in 10 µM $CaCl_2$ set at different pH values (4, 7, 9 and 11) as solvent. The values were taken averaging three different runs to get the standard deviation. The width of the size distribution is indicated by PDI, which in case of a monomodal distribution is defined as PDI=$(\delta/Z_{avg})$^2, where $\delta$ is the width of the distribution and $Z_{avg}$ is the average diameter of the particle population.

Silica Core diameter distribution was evaluated by the TEM images. Sample were prepared by drop casting and drying of a diluted water solution of PluS-X (1:25) on a Formvar film supported on standard 3.05 mm copper grid (400 mesh). The images were taken with a Philips CM 100 TEM operating at 80 KV. Silica core size distribution was calculated analyzing with ImageJ software TEM images containing a set of several hundreds of nanoparticles. The average diameter of the silica core was calculated fitting the profile of the histogram with a gaussian function.

*Calcite crystal growth* Calcium chloride dehydrate was purchased from Fluka, magnesium chloride hexahydrate from Sigma Aldrich and anhydrous ammonium carbonate from Acros Organics. All reagents were ACS grade and used as purchased. $CaCO_3$ crystals were grown by vapor diffusion method. Briefly, $(NH_4)_2CO_3$ vapor is allowed to diffuse into 750 µL of a 10 mM $CaCl_2$ solution containing the additives into a closed container for 4 days. The obtained crystals are then rinsed 3 times with water, once with ethanol and air-dried.

*Overgrowth experiments* $CaCO_3$ crystals to be used as seeds were grown by vapor diffusion method ad described above (see crystal growth and characterization). Calcite seeds were then put in a solution containing 750 µL of a 10 mM $CaCl_2$ solution containing the additives in a closed container in the presence of $(NH_4)_2CO_3$. After 4 days, the crystals were washed 3 times with water, once with ethanol and air-dried.

*PluS-X desorption experiments* The amount of particles released by the crystals (wt.%) after washing three times with water and once with ethanol has been determined by quantifying via fluorescence spectroscopy (Fluoromax-4 Horiba equipped with 150 W Xenon lamp single monochromator for both excitation and emission ($\lambda$exc=530 nm, slit width=8x8, scan from 550 nm to 750 nm) the PluS-X released in a saturated $CaCO_3$ solution (1 mL), in which the crystals were submerged for 72 hours under shaking. Calcium content was measured using flame atomic absorption spectroscopy (Perkin-Elmer AAnalyst 100). PluS-X loading was calculated assuming that a PluS-X particle weights 1M Da. The wt.% of released PluS-X is Table S2.

*Crystal characterization* SEM images were collected using a Zeiss Gemini SEM Ultra Plus with an acceleration voltage of 1.00 KeV after coating the sample with 2 nm of PtPd.

In order to measure the PluS-X content, the samples were first treated with 5% v/v sodium hypochlorite to bleach the fluorescence of the PluS-X adsorbed on the surface, then dissolved in 0.8 mL 0.1 M citrate buffer at pH 4.5 and finally the emission spectra were collected using Fluoromax-4 Horiba equipped with 150 W Xenon lamp single monochromator for both excitation and emission ($\lambda_{exc}$=530 nm, slit width=8x8, scan from 550 nm to 750 nm). Calcium content was measured using flame atomic absorption spectroscopy (Perkin-Elmer AAnalyst 100). PluS-X loading was calculated assuming that a PluS-X particle weights 1M Da.

High resolution X-ray powder diffraction (HRPXRD) measurements were collected with a dedicated high-resolution powder diffraction synchrotron beamline (ID22 at the European Synchrotron Radiation Facility (ESRF), Grenoble, France) at a wavelength of 0.039 nm. Crystal lattice parameters were assessed by the Rietveld refinement method with GSAS software and the EXPGUI interface. [51]

High resolution scanning transmission electron microscopy (HRTEM), high-angle annular dark-field scanning transmission electron microscopy (HAADF-STEM) images and electron dispersive spectroscopy (EDS) maps were acquired using a monochromated and double corrected Titan Themis G2 60-300 (FEI / Thermo Fisher) operated at 200KeV and equipped with a DualX detector (Bruker). The quantitative analysis of the EDS maps was done using the Velox software. Thin sections for TEM analysis were prepared using FEI Strata 400S FIB.

Confocal microscope images were collected using an Upright Zeiss LSM 710 equipped with a laser diode (561 nm) and analysed using the software imageJ, making a sum of the slides.

ASSOCIATED CONTENT

**Supporting Information**. Experimental section and further characterization (PDF)

AUTHOR INFORMATION

**Corresponding Authors**

Giuseppe Falini; e-mail: giuseppe.falini@unibo.it
Boaz Pokroy; e-mail: bpokroy@tx.technion.ac.il

**Author Contributions**

The manuscript was written through contributions of all authors. All authors have given approval to the final version of the manuscript.

ACKNOWLEDGMENT

SEM imaging have been collected at the Center for Nanoscale Systems (CNS), a member of the National Nanotechnology Coordinated Infrastructure Network (NNCI), which is supported by the National Science Foundation under NSF award no. 1541959. CNS is part of Harvard

University. BP acknowledges funding from the European Research Council under the European Union's Seventh Framework Program (FP/2007-2013)/ERC Grant Agreement [number 336077]. We are indebted to Yaron Kauffman from the Technion for help in acquiring the TEM data. BP and IP thank the ID22 high-resolution powder diffraction synchrotron beamline at the European Synchrotron Radiation Facility (ESRF; Grenoble, France). GF and GM thank the Consorzio Interuniversitario per la Ricerca sulla Chimica dei Metalli nei Sistemi Biologici for the support.

REFERENCES

(1)     Lowenstam, H. A.; Weiner, S. *On Biomineralization*; Oxford University Press, 1989.

(2)     Weiner, S.; Addadi, L. Crystallization Pathways in Biomineralization. *Annu. Rev. Mater. Res.* **2011**, *41* (1), 21–40.

(3)     Weiner, S.; Addadi, L. Design Strategies in Mineralized Biological Materials. *J. Mater. Chem.* **1997**, *7* (5), 689–702.

(4)     Estroff, L. A. Introduction: Biomineralization. *Chem. Rev.* **2008**, *108* (11), 4329–4331.

(5)     Nudelman, F.; Sommerdijk, N. A. J. M. Biomineralization as an Inspiration for Materials Chemistry. *Angew. Chem. Int. Ed.* **2012**, *51* (27), 6582–6596.

(6)     Schmidt, I.; Wagermaier, W. Tailoring Calcium Carbonate to Serve as Optical Functional Material: Examples From Biology and Materials Science. *Adv. Mater. Interfaces* **2016**, *4* (1), 1600250–1600256.

(7)     Karaseva, O. N.; Lakshtanov, L. Z.; Okhrimenko, D. V.; Belova, D. A.; Generosi, J.; Stipp, S. L. S. Biopolymer Control on Calcite Precipitation. *Crystal Growth & Design* **2018**, *18* (5), 2972–2985.

(8)     Addadi, L.; Weiner, S. Interactions Between Acidic Proteins and Crystals: Stereochemical Requirements in Biomineralization. *Proc Natl Acad Sci USA* **1985**, *82* (12), 4110–4114.

(9)     Metzler, R. A.; Evans, J. S.; Killian, C. E.; Zhou, D.; Churchill, T. H.; Appathurai, N. P.; Coppersmith, S. N.; Gilbert, P. U. P. A. Nacre Protein Fragment Templates Lamellar Aragonite Growth. *J. Am. Chem. Soc.* **2010**, *132* (18), 6329–6334.

(10)    Weiner, S.; Addadi, L. Acidic Macromolecules of Mineralized Tissues: the Controllers of Crystal Formation. *Trends in Biochemical Sciences* **1991**, *16*, 252–256.

(11)    Mann, S.; Archibald, D. D.; Didymus, J. M.; Heywood, B. R.; Meldrum, F. C.; Wade, V. J. Biomineralization: Biomimetic Potential at the Inorganic-Organic Interface. *MRS Bull.* **1992**, *17* (10), 32–36.

(12)    Luisa Fdez-Gubieda, M.; Muela, A.; Alonso, J.; Garcia-Prieto, A.; Olivi, L.; Fernandez-Pacheco, R.; Manuel Barandiaran, J. Magnetite Biomineralization in Magnetospirillum Gryphiswaldense: Time-Resolved Magnetic and Structural Studies. *ACS Nano* **2013**, *7* (4), 3297–3305.

(13)    Nudelman, F.; Gotliv, B.-A.; Addadi, L.; Weiner, S. Mollusk Shell Formation: Mapping the Distribution of Organic Matrix Components Underlying a Single Aragonitic Tablet in Nacre. *Journal of Structural Biology* **2006**, *153* (2), 176–187.

(14)     Albeck, S.; Aizenberg, J.; Addadi, L.; Weiner, S. Interactions of Various Skeletal
         Intracrystalline Components with Calcite Crystals. *J. Am. Chem. Soc.* **2002**, *115* (25),
         11691–11697.

(15)     Falini, G.; Albeck, S.; Weiner, S.; Addadi, L. Control of Aragonite or Calcite
         Polymorphism by Mollusk Shell Macromolecules. *Science* **1996**, *271* (5245), 67–69.

(16)     Berman, A.; Addadi, L.; Weiner, S. Interactions of Sea-Urchin Skeleton
         Macromolecules with Growing Calcite Crystals— a Study of Intracrystalline Proteins.
         *Nature* **1988**, *331* (6156), 546–548.

(17)     Heberling, F.; Trainor, T. P.; Luetzenkirchen, J.; Eng, P.; Denecke, M. A.; Bosbach, D.
         Structure and Reactivity of the Calcite-Water Interface. *Journal of Colloid and Interface
         Science* **2011**, *354* (2), 843–857.

(18)     Gotliv, B.-A.; Kessler, N.; Sumerel, J. L.; Morse, D. E.; Tuross, N.; Addadi, L.; Weiner,
         S. Asprich: a Novel Aspartic Acid-Rich Protein Family From the Prismatic Shell Matrix
         of the Bivalve Atrina Rigida. *ChemBioChem* **2005**, *6* (2), 304–314.

(19)     Evans, J. S. "Tuning in" to Mollusk Shell Nacre- and Prismatic-Associated Protein
         Terminal Sequences. Implications for Biomineralization and the Construction of High
         Performance Inorganic−Organic Composites. *Chem. Rev.* **2008**, *108* (11), 4455–4462.

(20)     Aizenberg, J.; Black, A. J.; Whitesides, G. M. Control of Crystal Nucleation by
         Patterned Self-Assembled Monolayers. *Nature* **1999**, *398* (6727), 495–498.

(21)     Addadi, L.; Moradian, J.; Shay, E.; Maroudas, N. G.; Weiner, S. A Chemical Model for
         the Cooperation of Sulfates and Carboxylates in Calcite Crystal Nucleation: Relevance
         to Biomineralization. *Proc Natl Acad Sci USA* **1987**, *84* (9), 2732–2736.

(22)     Green, D. C.; Ihli, J.; Kim, Y.-Y.; Chong, S. Y.; Lee, P. A.; Empson, C. J.; Meldrum, F.
         C. Rapid Screening of Calcium Carbonate Precipitation in the Presence of Amino Acids:
         Kinetics, Structure, and Composition. *Crystal Growth & Design* **2016**, *16* (9), 5174–
         5183.

(23)     Borukhin, S.; Bloch, L.; Radlauer, T.; Hill, A. H.; Fitch, A. N.; Pokroy, B. Screening the
         Incorporation of Amino Acids Into an Inorganic Crystalline Host: the Case of Calcite.
         *Adv. Funct. Mater. 22* (20), 4216–4224.

(24)     Magnabosco, G.; Giosia, M. D.; Polishchuk, I.; Weber, E.; Fermani, S.; Bottoni, A.;
         Zerbetto, F.; Pelicci, P. G.; Pokroy, B.; Rapino, S.; Falini, G.; Calvaresi, M. Calcite
         Single Crystals as Hosts for Atomic-Scale Entrapment and Slow Release of Drugs. *Adv.
         Healthcare Mater.* **2015**, *4* (10), 1510–1516.

(25)     Ukrainczyk, M.; Gredičak, M.; Jerić, I.; Kralj, D. Interactions of Salicylic Acid
         Derivatives with Calcite Crystals. *Journal of Colloid and Interface Science* **2012**, *365*
         (1), 296–307.

(26)     Kim, Y.-Y.; Fielding, L. A.; Kulak, A. N.; Nahi, O.; Mercer, W.; Jones, E. R.; Armes, S.
         P.; Meldrum, F. C. Influence of the Structure of Block Copolymer Nanoparticles on the
         Growth of Calcium Carbonate. *Chem. Mater.* **2018**, *30* (20), 7091–7099.

(27)     Ning, Y.; Whitaker, D. J.; Mable, C. J.; Derry, M. J.; Penfold, N. J. W.; Kulak, A. N.;
         Green, D. C.; Meldrum, F. C.; Armes, S. P. Anionic Block Copolymer Vesicles Act as
         Trojan Horses to Enable Efficient Occlusion of Guest Species Into Host Calcite Crystals.
         *Chem. Sci.* **2018**, *17*, 5218–6.

(28)     Kim, Y.-Y.; Freeman, C. L.; Gong, X.; Levenstein, M. A.; Wang, Y.; Kulak, A.;
         Anduix-Canto, C.; Lee, P. A.; Li, S.; Chen, L.; Christenson, H. K.; Meldrum, F. C. The

Effect of Additives on the Early Stages of Growth of Calcite Single Crystals. *Angew. Chem. Int. Ed. Engl.* **2017**, *56* (39), 11885–11890.

(29)   Kim, Y.-Y.; Semsarilar, M.; Carloni, J. D.; Cho, K. R.; Kulak, A. N.; Polishchuk, I.; Hendley, C. T., IV; Smeets, P. J. M.; Fielding, L. A.; Pokroy, B.; Tang, C. C.; Estroff, L. A.; Baker, S. P.; Armes, S. P.; Meldrum, F. C. Structure and Properties of Nanocomposites Formed by the Occlusion of Block Copolymer Worms and Vesicles Within Calcite Crystals. *Adv. Funct. Mater.* **2016**, *26* (9), 1382–1392.

(30)   Hanisch, A.; Yang, P.; Kulak, A. N.; Fielding, L. A.; Meldrum, F. C.; Armes, S. P. Phosphonic Acid-Functionalized Diblock Copolymer Nano-Objects via Polymerization-Induced Self-Assembly: Synthesis, Characterization, and Occlusion Into Calcite Crystals. *Macromolecules* **2016**, *49* (1), 192–204.

(31)   Kulak, A. N.; Yang, P.; Kim, Y.-Y.; Armes, S. P.; Meldrum, F. C. Colouring Crystals with Inorganic Nanoparticles. *Chemical Communications* **2014**, *50* (1), 67–69.

(32)   Ning, Y.; Fielding, L. A.; Doncom, K. E. B.; Penfold, N. J. W.; Kulak, A. N.; Matsuoka, H.; Armes, S. P. Incorporating Diblock Copolymer Nanoparticles Into Calcite Crystals: Do Anionic Carboxylate Groups Alone Ensure Efficient Occlusion? *ACS Macro Lett.* **2016**, *5* (3), 311–315.

(33)   Ning, Y.; Fielding, L. A.; Ratcliffe, L. P. D.; Wang, Y.-W.; Meldrum, F. C.; Armes, S. P. Occlusion of Sulfate-Based Diblock Copolymer Nanoparticles Within Calcite: Effect of Varying the Surface Density of Anionic Stabilizer Chains. *J. Am. Chem. Soc.* **2016**, *138* (36), 11734–11742.

(34)   Elhadj, S.; De Yoreo, J. J.; Hoyer, J. R.; Dove, P. M. Role of Molecular Charge and Hydrophilicity in Regulating the Kinetics of Crystal Growth. *Proc Natl Acad Sci USA* **2006**, *103* (51), 19237–19242.

(35)   Zuccarello, L.; Rampazzo, E.; Petrizza, L.; Prodi, L.; Satriano, C. The Influence of Fluorescent Silica Nanoparticle Surface Chemistry on the Energy Transfer Processes with Lipid Bilayers. *RSC Advances* **2016**, *6* (58), 52674–52682.

(36)   Wegner, G.; Demir, M. M.; Faatz, M.; Gorna, K.; Muñoz-Espí, R.; Guillemet, B.; Groehn, F. Polymers and Inorganics: a Happy Marriage? *Macromolecular Research* **2007**, *15* (2), 95–99.

(37)   Li, C.; Qi, L. Bioinspired Fabrication of 3D Ordered Macroporous Single Crystals of Calcite From a Transient Amorphous Phase. *Angew. Chem. Int. Ed. Engl.* **2008**, *47* (13), 2388–2393.

(38)   Lu, C.; Qi, L.; Cong, H.; Wang, X.; Yang, J.; Yang, L.; Zhang, D.; Ma, J.; Cao, W. Synthesis of Calcite Single Crystals with Porous Surface by Templating of Polymer Latex Particles. *Chem. Mater.* **2005**, *17* (20), 5218–5224.

(39)   Kim, Y.-Y.; Ribeiro, L.; Maillot, F.; Ward, O.; Eichhorn, S. J.; Meldrum, F. C. Bio-Inspired Synthesis and Mechanical Properties of Calcite–Polymer Particle Composites. *Adv. Mater.* **2010**, *22* (18), 2082–2086.

(40)   Ning, Y.; Han, L.; Derry, M. J.; Meldrum, F. C.; Armes, S. P. Model Anionic Block Copolymer Vesicles Provide Important Design Rules for Efficient Nanoparticle Occlusion Within Calcite. *J. Am. Chem. Soc.* **2019**, jacs.8b12507.

(41)   Burns, A.; Ow, H.; Wiesner, U. Fluorescent Core–Shell Silica Nanoparticles: Towards "Lab on a Particle" Architectures for Nanobiotechnology. *Chemical Society Reviews* **2006**, *35* (11), 1028–1042.

(42)  Helle, M.; Rampazzo, E.; Monchanin, M.; Marchal, F.; Guillemin, F.; Bonacchi, S.; Salis, F.; Prodi, L.; Bezdetnaya, L. Surface Chemistry Architecture of Silica Nanoparticles Determine the Efficiency of in Vivo Fluorescence Lymph Node Mapping. *ACS Nano* **2013**, *7* (10), 8645–8657.

(43)  Rampazzo, E.; Voltan, R.; Petrizza, L.; Zaccheroni, N.; Prodi, L.; Casciano, F.; Zauli, G.; Secchiero, P. Proper Design of Silica Nanoparticles Combines High Brightness, Lack of Cytotoxicity and Efficient Cell Endocytosis. *Nanoscale* **2013**, *5* (17), 7897–7905.

(44)  Hendley, C. T.; Fielding, L. A.; Jones, E. R.; Ryan, A. J.; Armes, S. P.; Estroff, L. A. Mechanistic Insights Into Diblock Copolymer Nanoparticle-Crystal Interactions Revealed via in Situ Atomic Force Microscopy. *J. Am. Chem. Soc.* **2018**, *140* (25), 7936–7945.

(45)  Mahrouqi, Al, D.; Vinogradov, J.; Jackson, M. D. Zeta Potential of Artificial and Natural Calcite in Aqueous Solution. *Advances in Colloid and Interface Science* **2017**, *240*, 60–76.

(46)  Stipp, S. L. S. Toward a Conceptual Model of the Calcite Surface: Hydration, Hydrolysis, and Surface Potential. *Geochimica et Cosmochimica Acta* **1999**, *63* (19-20), 3121–3131.

(47)  Rampazzo, E.; Bonacchi, S.; Juris, R.; Montalti, M.; Genovese, D.; Zaccheroni, N.; Prodi, L.; Rambaldi, D. C.; Zattoni, A.; Reschiglian, P. Energy Transfer From Silica Core−Surfactant Shell Nanoparticles to Hosted Molecular Fluorophores. *J. Phys. Chem. B* **2010**, *114* (45), 14605–14613.

(48)  Genovese, D.; Bonacchi, S.; Juris, R.; Montalti, M.; Prodi, L.; Rampazzo, E.; Zaccheroni, N. Prevention of Self-Quenching in Fluorescent Silica Nanoparticles by Efficient Energy Transfer. *Angew. Chem.* **2013**, *125* (23), 6081–6084.

(49)  Rampazzo, E.; Bonacchi, S.; Genovese, D.; Juris, R.; Montalti, M.; Paterlini, V.; Zaccheroni, N.; Dumas-Verdes, C.; Clavier, G.; Méallet-Renault, R.; Prodi, L. Pluronic-Silica (PluS) Nanoparticles Doped with Multiple Dyes Featuring Complete Energy Transfer. *J. Phys. Chem. C* **2014**, *118* (17), 9261–9267.

(50)  Genovese, D.; Rampazzo, E.; Bonacchi, S.; Montalti, M.; Zaccheroni, N.; Prodi, L. Energy Transfer Processes in Dye-Doped Nanostructures Yield Cooperative and Versatile Fluorescent Probes. *Nanoscale* **2014**, *6* (6), 3022–3036.

(51)  Toby, B. H.; Dreele, Von, R. B.; IUCr. GSAS-II: the Genesis of a Modern Open-Source All Purpose Crystallography Software Package. *Journal of Applied Crystallography* **2013**, *46* (2), 544–549.

# TABLE OF CONTENT

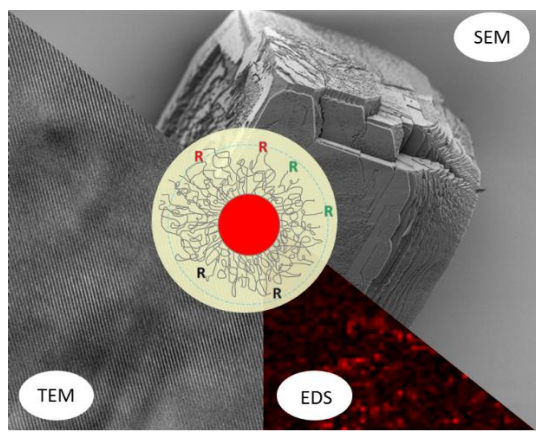